\documentclass[twocolumn,superscriptaddress,showpacs,floatfix,prc]{revtex4}
\usepackage{graphicx}

\begin{document}

\title{Pentaquark baryon production at the Relativistic Heavy Ion Collider}
\author{Lie-Wen Chen}
\thanks{On leave from Department of Physics, Shanghai Jiao Tong University,
Shanghai 200030, China}
\affiliation{Cyclotron Institute and Physics Department, Texas A\&M University, College
Station, Texas 77843-3366}
\author{V. Greco}
\affiliation{Cyclotron Institute and Physics Department, Texas A\&M University, College
Station, Texas 77843-3366}
\author{C. M. Ko}
\affiliation{Cyclotron Institute and Physics Department, Texas A\&M University, College
Station, Texas 77843-3366}
\author{S. H. Lee}
\thanks{On leave from Institute of Physics and Applied Physics and
Department of Physics, Yonsei University, Seoul 120-749, Korea}
\affiliation{Cyclotron Institute and Physics Department, Texas A\&M University, College
Station, Texas 77843-3366}
\author{W. Liu}
\affiliation{Cyclotron Institute and Physics Department, Texas A\&M University, College
Station, Texas 77843-3366}
\date{\today}

\begin{abstract}
Production of pentaquark $\Theta^+$ baryons in central relativistic heavy
ion collisions is studied in a kinetic model. Assuming that a quark-gluon
plasma is produced in the collisions, we first determine the number of $%
\Theta^+$ produced from the quark-gluon plasma using a parton coalescence
model, and then take into consideration its production and absorption in
subsequent hadronic matter via the reactions $KN\leftrightarrow\Theta$, $%
KN\leftrightarrow\pi\Theta$, and $\pi N\leftrightarrow\bar K\Theta$. We find
that although the final $\Theta^+$ number is affected by hadronic
interactions, it remains sensitive to the initial number of $\Theta^+$
produced from the quark-gluon plasma, particularly in the case of a small $%
\Theta^+$ width as imposed by the $K^+N$ and $K^+d$ scattering data. Because
of small baryon chemical potential in the hot dense matter produced in these
collisions, the number of produced anti-$\Theta$ is only slightly smaller
than that of $\Theta^+$.
\end{abstract}

\maketitle

\section{introduction}

In recent experiments on nuclear reactions induced by photons \cite%
{nakano,stepanyan,kubarovsky,barth,airapetian}, kaons \cite{barmin}, and
neutrinos \cite{asratyan}, production of baryons consisting of five quarks $%
uudd\bar{s}$ has been inferred from the invariant mass spectrum of $K^{+}n$
or $K^{0}p$. The extracted mass is around $1.536$ GeV with a width in the
range $20$-$25$ MeV except Ref.\cite{barmin} which gives a width of $9$ MeV.
All these widths reflect the resolution in the experiments, so the actual
width of $\Theta ^{+}$ is expected to be smaller. The observed properties of
this pentaquark baryon are consistent with those of the $\Theta ^{+}$ baryon
with spin $J=1/2$, isospin $I=0$, and strangeness $S=+1$ that was originally
predicted by the chiral soliton model \cite{diakonov} and recently studied
using the Skyrme model \cite%
{prasz,polyakov,walliser,jennings,borisyuk,itzhaki}, the constituent quark
model \cite{riska,lipkin,jaffe}, the chiral quark model \cite{hosaka,glozman}%
, the QCD sum rules \cite{zhu,matheus,sugiyama}, and the lattice QCD \cite%
{sasaki,ciskor}. Studies of the $\Theta ^{+}$ production mechanism in these
reactions have also been carried out \cite{liu,liu1,oh,nam}. Since both kaon
and nucleon numbers are not insignificant in the hadronic matter formed in
relativistic heavy ion collisions, the $\Theta ^{+}$ may also be produced in
these collisions. Using a statistical model, which assumes that the $\Theta
^{+}$ is in chemical equilibrium with other hadrons, Randrup \cite{randrup}
has estimated its abundance and finds that there is about one $\Theta ^{+}$
per unit rapidity in central Au+Au collisions at $\sqrt{s_{NN}}=200$ GeV
available from the Relativistic Heavy Ion Collider (RHIC). As the
quark-gluon plasma is expected to be formed in the initial stage of heavy
ion collisions at RHIC, and it was suggested that formation of the
quark-gluon plasma would enhance the production of hadrons consisting of
strange quarks \cite{rafelski}, it is of interest to know if this is also
the case for $\Theta ^{+}$ production and to determine its production from
the quark-gluon plasma relative to that from the hadronic matter.

In this Letter, $\Theta^+$ production in central heavy ion collisions at RHIC
is studied in a kinetic model that starts from the final stage of the
quark-gluon plasma, goes through a mixed phase of quark-gluon and hadronic
matters, and finally undergoes hadronic expansion. The production of $%
\Theta^+$ from the quark-gluon plasma is modeled by the coalescence model,
which has been shown to describe quite well not only the particle yields and
their ratios \cite{alcor} but also their transverse momentum spectra \cite%
{greco,hwa,fries}. Production and absorption of $\Theta^+$ in the hadronic
matter are taken into account via the reactions $KN\leftrightarrow\Theta$, $%
KN\leftrightarrow\pi\Theta$, and $\pi N\leftrightarrow\bar K\Theta$. We find
that the number of $\Theta^+$ produced from the quark-gluon plasma is
appreciable and the final $\Theta^+$ number after the hadronic phase remains
sensitive to the initial number of $\Theta^+$ produced from the quark-gluon
plasma. Furthermore, a slightly smaller number of its antiparticle $%
\bar\Theta^-$ is also expected to be produced at RHIC as a result of small
baryon chemical potential in the matter produced from these collisions. We
note that the quark coalescence model has also been used recently to study
in heavy ion collisions at RHIC the elliptic flow of $\Theta^+$ \cite{nonaka}%
, i.e., the azimuthal anisotropy of their momentum distribution in the plane
perpendicular to the beam directions.

\section{Collision dynamics at RHIC}

To model the dynamics of central relativistic heavy ion collisions after the
end of the quark-gluon plasma phase, we use the boost invariant picture of
Bjorken \cite{bjorken} augmented with accelerated transverse expansion.
Specifically, the volume of produced fireball is taken to evolve with the
proper time according to \cite{ko} 
\begin{equation}
V(\tau )=\pi \left[ R_{\mathrm{C}}+v_{\mathrm{C}}(\tau -\tau _{\mathrm{C}%
})+a/2(\tau -\tau _{\mathrm{C}})^{2}\right] ^{2}\tau c,
\end{equation}%
where $R_{\mathrm{C}}=8$ fm and $\tau _{\mathrm{C}}=5$ fm/$c$ are final
transverse and longitudinal sizes of the quark-gluon plasma, corresponding
to a volume of $1,006$ fm$^{3}$ in central Au+Au collisions at $\sqrt{s_{NN}}%
=200$ GeV. Taking the critical temperature to be $T_{\mathrm{C}}=175$ MeV
and a transverse flow velocity $v_{\mathrm{C}}=0.4c$, the total transverse
energy of quarks and gluons in the midrapidity ($|y|\leq 0.5$) is then about 
$1,067$ GeV, if we take quarks and gluons to be massive with $m_{g}=500$
MeV, $m_{u}=m_{d}=300$ MeV and $m_{s}=475$ MeV in order to take into account
the nonperturbative effects of QCD near critical temperature \cite{levai},
and to have chemical potentials $\mu _{b}=10$ MeV and $\mu _{s}=0$ to
account for observed final antibaryon to baryon ratio at RHIC. The above
total transverse energy includes a bag energy of about $133.8$ GeV based on
a bag constant of about $133$ MeV/fm$^{3}$, which is determined from the
pressure difference between the quark-gluon plasma and the hadronic matter
at $T_{C}$. Requiring that final hadronic matter freezes out at temperature $%
T_{\mathrm{F}}=125$ MeV and has a transverse flow velocity of $0.65c$, as
extracted from measured hadron spectra in these collisions, then leads to a
freeze out volume $V(\tau _{\mathrm{F}})\approx 11,322$ fm$^{3}$ if we
assume that the fireball expands isentropically. The resulting total
transverse energy of final hadronic matter is $788$ GeV and is comparable to
that of midrapidity hadrons measured in experiments \cite{adcox}. It is,
however, less than the initial transverse energy of midrapidity quarks and
gluons. This is consistent with the reduction of midrapidity hadrons and
their total transverse energy as a result of hadronic rescatterings seen in
studies based on the transport model \cite{ampt}. Because of the small
pressure near phase transition and in hadronic matter \cite{karsh},
acceleration in the transverse expansion is chosen to have a small value $%
a=0.02$ $c^{2}$/fm in order to obtain a lifetime of the expanding matter
comparable to that from the transport model.

\begin{figure}[th]
\includegraphics[scale=1.2]{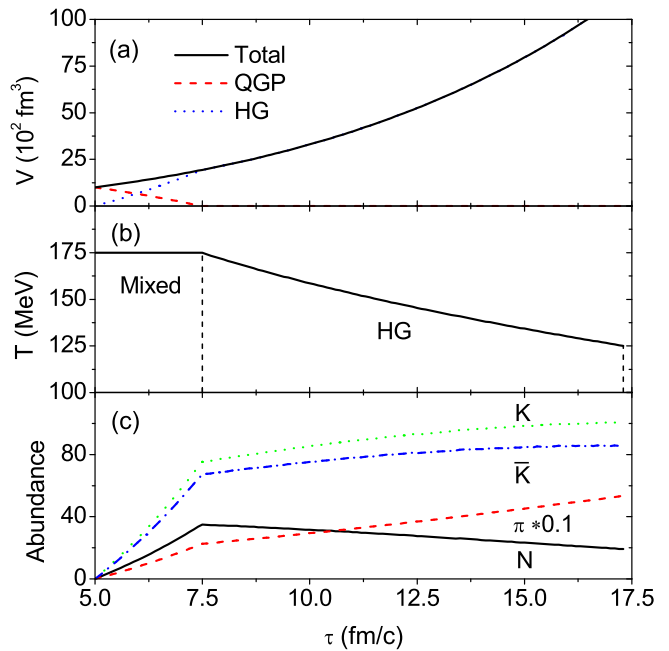}
\caption{(Color online) Time evolution of the volume (a), temperature (b),
and abundance (c) of mid-rapidity particles in Au+Au collisions at $\protect%
\sqrt{s_{NN}}=200$ GeV.}
\label{voltem}
\end{figure}

As the quark-gluon plasma expands, it gradually converts to hadrons leading
to a mixed phase of quark-gluon plasma and hadronic matter at constant
temperature $T_{\mathrm{C}}$. The fraction of hadronic matter $f_{\mathrm{H}%
}(\tau )$ can be determined by requiring that the total entropy $S_{\mathrm{%
tot}}$ of the fireball remains constant during the phase transition \cite%
{kajantie}, i.e., 
\begin{equation}
f_{\mathrm{H}}(\tau )s_{\mathrm{H}}(T_{\mathrm{C}})+(1-f_{\mathrm{H}}(\tau
))s_{\mathrm{QGP}}(T_{\mathrm{C}})=\frac{S_{\mathrm{tot}}}{V(\tau)},
\end{equation}
where $s_{\mathrm{H}}$ and $s_{\mathrm{QGP}}$ are the entropy densities of
hadronic matter and quark-gluon plasma, respectively, which we treat as
noninteracting ideal gases with finite transverse flow. The mixed phase ends
at time $\tau _{\mathrm{H}}\approx 7.5$ fm/$c$ when $f_{\mathrm{H}}(\tau _{%
\mathrm{H}})$ reaches one, and the resulting pure hadronic matter is also
assumed to expand isentropically until $\tau _{\mathrm{F}}\approx 17.3$ fm/$%
c $ when its temperature drops to $T_{F}=125$ MeV. In panels (a) and (b) of
Fig.\ref{voltem}, we show the resulting time evolution of the volume and
temperature of mid-rapidity particles in Au+Au collisions at $\sqrt{s_{NN}}%
=200$ GeV.

For normal hadrons such as pions, kaons, and nucleons, we take them to be in
chemical equilibrium with baryon chemical potential $\mu_{B}=30$ \textrm{MeV}%
, charge chemical potential $\mu_{Q}=0$ \textrm{MeV}, and strangeness
chemical potential $\mu_{S}=10$ \textrm{MeV}, similar to those of the
quark-gluon plasma. Since the chemical potentials vary weakly with the
temperature of an isentropically expanding matter in heavy ion collisions at
RHIC energies \cite{rapp}, we neglect their time dependence. Evaluating the
densities of these hadrons using the relativistic Boltzmann distributions
with inclusion of transverse flow, the time evolution of their abundance is
shown in panel (c) of Fig.\ref{voltem}. We note that the final numbers of
pions, kaons, and nucleons at freeze out including contributions from decays
of resonances are comparable to those measured in experiments.

\section{$\Theta^+$ production from the QGP}

The number of $\Theta^{+}$ baryons that are produced from the quark-gluon
plasma can be estimated using the coalescence model \cite{greco}, i.e.,
given by the product of a statistical factor $g_\Theta$, which denotes the
probability of combining $uudd\bar{s}$ quarks into a color neutral, spin $%
1/2 $, and isospin $0$ hadronic state, and the overlap of the quark
phase-space distribution function $f_{q}(x_{i},p_{i})$ with the Wigner
distribution function $f_{\Theta }$ of $\Theta ^{+}$. The latter corresponds
to the probability of converting the above hadronic state into $\Theta^+$
and depends on the quark spatial wave functions in the $\Theta^+$.
Explicitly, the $\Theta^+$ number is expressed as 
\begin{equation}  \label{coal}
N_{\Theta }=g_{\Theta }\int \prod_{i=1}^{5}\frac{p_{i}\cdot d\sigma _{i}d^{3}%
\mathbf{p}_{i}}{(2\pi )^{3}E_{i}}f_{q}(x_{i},p_{i})f_{\Theta
}(x_{1}..x_{5};p_{1}..p_{5}),
\end{equation}%
with $d\sigma $ denoting an element of a space-like hypersurface. Since the
coalescence model can be viewed as formation of bound states from
interacting particles with energy mismatch balanced by other particles in
the system, neglecting such off-shell effects is reasonable if the binding
energy is not large as in the case of $\Theta ^{+}$ and/or if the production
process is fast compared to the inverse of the binding energy. We note that
the $\Theta^{+}$ is produced from the quark-gluon plasma during its
hadronization and thus exists only in the resulting hadronic matter.

To determine the statistical factor $g_\Theta$, we note that the quark wave
function of $\Theta^+$ in the color-spin-isospin space can be expressed as a
linear combination of all possible orthogonal flavor, color, and spin basis
states, with coefficients depending on the quark model used for $\Theta^+$,
e. g., see \cite{carlson}. Since these coefficients are normalized to one,
the probability for the five quarks and antiquark $uudd\bar s$ to form a
hadron with quantum numbers corresponding to a spin up $\Theta^+$ is simply
given by the probability of finding these quarks in any one of these
color-spin-isopin basis states, i.e., $1/3^{5}\times 1/2^{5}=1/7776$.
Including also the possibility of forming a spin-down $\Theta^+$ doubles the
probability. As a result, the probability factor is $g_\Theta=1/3888$.

For the phase-space distribution function of quarks, we assume that they are
uniformly distributed in the transverse plane and their momentum
distributions are relativistic Boltzmannian in the transverse direction but
uniform in rapidity along the longitudinal direction. Imposing also the
Bjorken correlation of equal spatial ($\eta$) and momentum ($y$) rapidities,
then the quark momentum distribution per unit rapidity is 
\begin{equation}
f_q(\eta,y,\mathbf{p}_\bot)=g_q\delta(\eta-y) \exp\left(\frac{-\sqrt{\mathbf{%
p}_\bot^2+m_q^2}}{T}\right),
\end{equation}
where $g_q=6$ is the color-spin degeneracy of a quark.

The Wigner distribution function of $\Theta ^{+}$ depends on its internal
quark wave functions. For simplicity, we take it to be Gaussian in space and
momentum, i.e.,
\begin{equation}
f_{\Theta }(x;p)=8^{4}\exp \left( -\sum_{i=1}^{4}\frac{\mathbf{y}_{i}^{2}}{%
\sigma _{i}^{2}}-\sum_{i=1}^{4}\mathbf{k}_{i}^{2}\sigma _{i}^{2}\right) ,
\end{equation}%
where $\mathbf{y}_{i}$ and $\mathbf{k}_{i}$ are the relative spatial and
momentum coordinates of the five quarks in $\Theta ^{+}$, and are related to 
$\mathbf{x}_{i}$ and $\mathbf{p}_{i}$ by the normal Jacobian transformation,
i.e., 
\begin{equation}
\mathbf{y}_{i}=\left( 1+\frac{1}{i}\right) ^{-1/2}\left( \frac{%
\sum_{j=1}^{i}m_{j}\mathbf{x}_{j}}{\sum_{j=1}^{i}m_{j}}-\mathbf{x}%
_{i+1}\right)
\end{equation}%
and similarly for the momentum space coordinates. The width parameter $%
\sigma _{i}^{2}$ for the $i$-th relative coordinate is defined as $\sigma
_{i}^{2}=\left( \mu _{i}\omega \right) ^{-1}$ with 
\begin{equation}
\mu _{i}=\frac{1+\frac{1}{i}}{\frac{1}{m_{i+1}}+\frac{1}{\sum_{j=1}^{i}{m_{j}%
}}}.
\end{equation}

Carrying out the spatial integrations in Eq.(\ref{coal}), the number of $%
\Theta ^{+}$ produced from the coalescence of $uudd\bar{s}$ quarks is then 
\begin{eqnarray}
N_{\Theta } &\simeq &g_{\Theta }N_{u}^{2}N_{d}^{2}N_{\bar{s}}\left( {\frac{4%
}{5}}{\frac{m_{q}}{m_{\bar{s}}}}+{\frac{1}{5}}\right) ^{\frac{3}{2}}\left( 
\frac{(4\pi \sigma ^{2})^{\frac{3}{2}}}{V}\right) ^{4}  \nonumber \\
&&\times \frac{\int {\ \prod_{i=1}^{5}{dm_{i\perp }\,m_{i\perp
}^{2}\,e^{-m_{i\perp }/T}\prod_{j=1}^{4}e^{-k_{j\perp }^{^{\prime }2}\sigma
_{j}^{2}}}}}{\prod_{i=1}^{5}e^{-m_{i}/T}(m_{i}^{2}T+2m_{i}T^{2}+2T^{3})}.
\label{number}
\end{eqnarray}%
In the above, $\mathbf{k}_{j\perp }^{^{\prime }}$ are the four relative
momenta defined previously but determined in the center-of-mass of formed
pentaquark, and $m_{i\perp }=\sqrt{m_{i}^{2}+\mathbf{p}_{\perp }^{2}}$ are
transverse masses of the quarks. The numbers of $u$, $d$, and $\bar{s}$
quarks in the quark-gluon plasma with rapidities $|y|\leq 0.5$ are denoted,
respectively, by $N_{u}$, $N_{d}$, and $N_{\bar{s}}$; and $\sigma $ is the
width parameter calculated according to the mass $m_{q}$ of the light quark.
At $\tau _{\mathrm{C}}$ their values are $N_{u}=N_{d}\simeq 245$ and $N_{%
\bar{s}}\simeq 149$, if we take into account the effect of gluons by
converting them to quarks according to the quark flavor composition in the
quark-gluon plasma as in Ref.\cite{greco}. The width parameter $\sigma $ in
the $\Theta ^{+}$ Wigner distribution function is related to the size of $%
\Theta ^{+}$. Since the latter is not known empirically, we thus choose the
value of $\sigma $ to fit the proton root-mean-square radius with the
harmonic oscillator wave functions, i.e., $\sigma =\langle r_{\mathrm{p}%
}^{2}\rangle ^{1/2}\approx 0.86$ fm \cite{simon}, corresponding to a $\Theta
^{+}$ root-mean-square radius 
\begin{equation}
\langle r_{\Theta }^{2}\rangle ^{1/2}=\sqrt{\frac{6}{5}}\sigma \left( 1-%
\frac{1-\frac{m_{q}}{m_{s}}}{4+\frac{m_{q}}{m_{s}}}\right) ^{1/2}\approx 0.9~%
\mathrm{fm}.
\end{equation}

Evaluating numerically the transverse mass integrals in Eq.(\ref{number})
using the Monte Carlo method introduced in Ref.\cite{greco}, we find that
the number of $\Theta ^{+}$ produced in one unit of rapidity is about $0.19$%
. Compared to predictions from the statistical model, this number is smaller
than that estimated in Ref.\cite{randrup} by about a factor five and that
from a more recent analysis \cite{letessier} by about a factor two. The
coalescence model also allows us to evaluate the number $N_{Y}$ of neutral
hyperons $\Lambda $ and $\Sigma ^{0}$ as well as those from decays of $\Xi
^{0}$ and $\Xi ^{-}$ that are produced from the quark-gluon plasma. The
resulting number is about $10$. Although this is about a factor of two
smaller than that measured in Au+Au collisions at $\sqrt{s}=130$ GeV \cite%
{star}, including contributions from resonance decays and production in the
hadronic matter is expected to bring it closer to the experimental data. In
deriving Eq.(\ref{number}), we have neglected the effect of transverse flow
on the quark momentum distributions. Including this effect makes the spatial
integrals in Eq.(\ref{coal}) more involved. Evaluating them by the Monte
Carlo method, we find that its effect on produced $\Theta ^{+}$ number is
small. It is interesting to note that the momentum integrals, i.e., the last
factor in Eq.(\ref{number}), can be evaluated analytically if the quark
momentum distributions are non-relativistic, leading to the expression $%
(1+2m_{q}T\sigma ^{2})^{-4}$. This gives a $\Theta ^{+}$ number of 0.64,
which is about a factor of $3.4$ larger than the relativistic case.

\section{Hadronic effects on $\Theta^+$ production}

The abundance of $\Theta^+$ can change during subsequent evolution of the
hadronic matter as a result of hadronic absorption and production. In
hadronic matter, $\Theta^+$ can be produced as a resonance from interactions
of $K$ and $N$. The cross section for this process is given by the
Breit-Wigner formula. It can also be produced from reactions like $%
KN\to\pi\Theta$ and $\pi N\to\bar K\Theta$. Cross sections for these
reactions can be estimated by considering $KN\to\pi\Theta$ as a $u-$channel
and $\pi N\to\bar K\Theta$ as an $s-$channel nucleon-pole diagram. Using the
coupling constant $g_{KN\Theta}\simeq 4.4$, determined from the width of $%
\Theta^+$ which we take to be $\Gamma_\Theta=20$ MeV, we have evaluated
these cross sections with empirical form factors. Details of this
calculation can be found in Ref.\cite{liu}. The $\Theta^+$ can be destroyed
in the hadronic matter either by decay, i.e., $\Theta\to KN$, or by the
inverse reactions $\pi\Theta\to KN$ and $\bar K\Theta\to\pi N$ with cross
sections related to those of production cross sections via the detailed
balance relations.

We have also evaluated the thermal average of $\Theta ^{+}$ production and
absorption cross sections. In the temperature range $125$-$175$ MeV of
interest here, their values for production cross sections are about $1$ mb
for $\langle \sigma _{KN\rightarrow \Theta }v\rangle $, $0.3$ mb for $%
\langle \sigma _{KN\rightarrow \pi \Theta }v\rangle $, and $1$-$3$ $\mu $b
for $\langle \sigma _{\pi N\rightarrow \bar{K}\Theta }v\rangle $, where $v$
is the relative velocity of two initial hadrons and $\langle \cdots \rangle $
denotes the average over their momentum distributions. For $\Theta ^{+}$
absorption cross sections, they are about $6$ mb for $\langle \sigma _{\pi
\Theta \rightarrow KN}v\rangle $ and $0.6$ mb for $\langle \sigma _{\bar{K}%
\Theta \rightarrow \pi N}v\rangle $. The cross sections are thus larger for $%
\Theta ^{+}$ production reactions induced by $K$ than $\pi $ and also for $%
\Theta ^{+}$ absorption than production reactions as a result of large $%
\Theta ^{+}$ mass.

In terms of thermal averaged cross sections and the densities of pions ($%
n_{\pi }$), kaons ($n_{K}$), antikaons ($n_{\bar{K}}$), and nucleons ($n_{N}$%
), the time evolution of $\Theta ^{+}$ abundance $N_{\Theta }$ is determined
by the kinetic equation \cite{xia,Ko01},%
\begin{eqnarray}
\frac{dN_{\Theta }}{d\tau } &=&R_{\mathrm{QGP}}(\tau )+\left\langle \sigma
_{\pi N\rightarrow \bar{K}\Theta }v\right\rangle n_{\pi }n_{N}V_{\mathrm{H}}
\nonumber \\
&&+(\left\langle \sigma _{KN\rightarrow \pi \Theta }v\right\rangle
+\left\langle \sigma _{KN\rightarrow \Theta }v\right\rangle )n_{K}n_{N}V_{%
\mathrm{H}}  \nonumber \\
&&-\Gamma _{\Theta }N_{\Theta }-\left\langle \sigma _{\bar{K}\Theta
\rightarrow \pi N}v\right\rangle n_{\bar{K}}N_{\Theta }  \nonumber \\
&&-\left\langle \sigma _{\pi \Theta \rightarrow KN}v\right\rangle n_{\pi
}N_{\Theta }.  \label{kinetic}
\end{eqnarray}%
Since hadronization of the quark-gluon plasma takes a finite time of $\tau _{%
\mathrm{H}}-\tau _{\mathrm{C}}\simeq 2.5$ fm/$c$, we expect $\Theta ^{+}$ to
be produced continuously from the quark-gluon plasma in the mixed phase,
with a rate proportional to the volume of the quark-gluon plasma. This is
included in the first term $R_{\mathrm{QGP}}(\tau )$ of above equation.

\begin{figure}[th]
\includegraphics[scale=1.0]{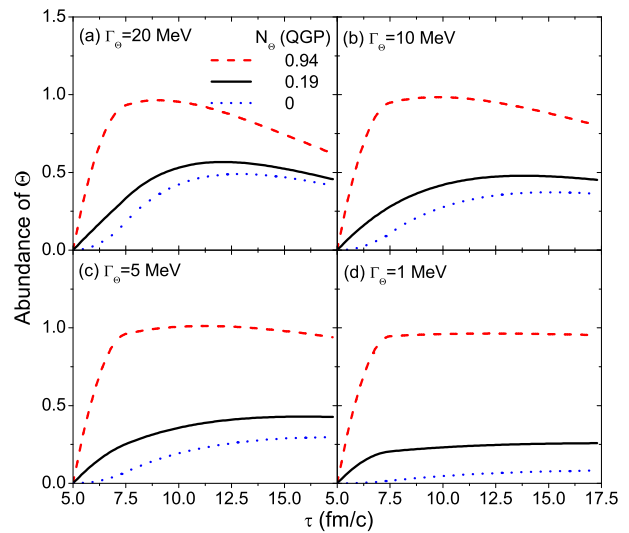}
\caption{(Color online) Time evolution of $\Theta ^{+}$ abundance in Au+Au
collisions at $\protect\sqrt{s_{NN}}=200$ GeV for different initial numbers
of $\Theta ^{+}$ produced from the quark-gluon plasma as well as different $%
\Theta ^{+}$ width.}
\label{mult}
\end{figure}

Using the time evolution of volume, temperature, and hadron abundance shown
in Fig.\ref{voltem}, the resulting time evolution of $\Theta ^{+}$ abundance
is shown in Fig.\ref{mult}(a). The solid curve is obtained with $0.19$
initial $\Theta ^{+}$ from the quark-gluon plasma as given by the
coalescence model discussed in the above. It is seen that the final $\Theta
^{+}$ number is enhanced to $0.46$. To see how the final $\Theta ^{+}$
number depends on its number produced from the quark-gluon plasma, we have
repeated the above calculation with different numbers of initial $\Theta ^{+}
$ baryons. As shown in Fig.\ref{mult}(a), the final $\Theta ^{+}$ number is
about $0.41$ and $0.62$, respectively, for $0$ (dotted curve) and $0.94$
(dashed curve) initial $\Theta ^{+}$ baryons, roughly corresponding to a
change of $30\%$ in the value of $\sigma $ or the size of $\Theta ^{+}$. Our
results demonstrate that although the final number of $\Theta ^{+}$ produced
in relativistic heavy ion collision is affected by hadronic interactions, it
is nonetheless sensitive to the initial number of $\Theta ^{+}$ produced
from the quark-gluon plasma.

This sensitivity would be even stronger if the width of $\Theta ^{+}$ is
smaller than that used here, i.e., $\Gamma _{\Theta }=20$ MeV. In this case,
not only the decay probability of $\Theta ^{+}$ is reduced but also the
hadronic cross sections become smaller as they are proportional to the
square of the coupling constant $g_{KN\Theta }$, which is proportional to $%
\Gamma _{\Theta }$. This is shown in Figs. \ref{mult}(b), (c), and (d) for
the abundance of $\Theta ^{+}$ using different $\Theta ^{+}$ widths of $10$, 
$5$, and $1$ MeV, respectively. The smaller width of $5$ MeV or less is
actually more consistent with available $K^{+}N$ and $K^{+}d$ data \cite%
{arndt,haiden,nuss,cahn,azimov}. With such a small $\Theta ^{+}$ width, the
hadronic effects become insignificant, and the final $\Theta ^{+}$ abundance
is essentially determined by production from the quark-gluon plasma.

On the other hand, the dependence of final $\Theta ^{+}$ number on its
initial number from the quark-gluon plasma disappears only if hadronic cross
sections are large. This happens when the latter is increased by about a
factor of $3$, corresponding to a $\Theta ^{+}$ width of $60$ MeV. Although
such a large $\Theta ^{+}$ in-medium width is unlikely due to the weak
coupling of $\Theta ^{+}$ to $KN$, the final $\Theta ^{+}$ number in this
case is about $0.35$ and is comparable to that from the statistical model
prediction of \cite{letessier}.

\section{discussions}

In the kinetic equation, Eq.(\ref{kinetic}), we have not considered
reactions involving more than three particles such as $\pi \Theta
\leftrightarrow \pi KN$. Taking its cross section the same as that for the
reaction $\pi \Theta \leftrightarrow KN$, Eq.(\ref{kinetic}) can be
generalized to include both $\Theta ^{+}$ production and annihilation. We
find that including this reaction affects the final $\Theta ^{+}$ abundance
by less than $10\%$. Also, medium effects on the $\Theta ^{+}$ properties
are not included. Because of the weak $\Theta ^{+}$ coupling to $KN$ and the
small hadron density after hadronization of the quark-gluon plasma, these
effects are not expected to be important.

Our result that the final number of exotic pentaquark $\Theta^+$ produced in
heavy ion collisions at RHIC is sensitive to its initial number produced
during the hadronization of the quark-gluon plasma is, however, based on the
assumption that $\Theta^+$ production from initial nucleon-nucleon
collisions is not significant compared to that from the hot dense matter
formed during the collisions. This is expected to be the case as string
fragmentation to multiquark-anti-multiquark pairs is suppressed compared to
fragmentation to diquark-anti-diquark pairs, from which normal three-quark
baryons are produced in high energy proton-proton collisions. Furthermore,
both the quark flavor structure and mass of $\Theta^+$, which belongs to the
antidecuplet of SU(3), are similar to those of the decuplet $\bar\Omega^+$ ($%
\bar s\bar s\bar s$), whose abundance in central relativistic heavy ion
collisions is known to be more than an order of magnitude larger than that
expected from initial nucleon-nucleon interactions \cite{star}.

Our study can be applied to production of anti-$\Theta^+$ in heavy ion
collisions at RHIC. With the light antiquark to quark ratio of $0.89$ and
antistrange to strange quark ratio of $1$ as a result of small quark baryon
chemical potential $\mu _{b}=10$ MeV, which gives an antiproton to proton
ratio of $(0.89)^{3}\simeq 0.7$ at midrapidity consistent with experimental
observations, the $\bar\Theta^-$ to $\Theta^+$ ratio is expected to be $%
(0.89)^{4}\simeq 0.63$. Relativistic heavy ion collisions at RHIC thus also
allows us to find the $\bar\Theta^-$, which is not likely to be produced in
either photo- or kaon-nucleus reactions.

\section{summary}

To summarize, we have used a kinetic model to study production of pentaquark 
$\Theta^+$ baryon in central relativistic heavy ion collisions by including
contributions both from the quark-gluon plasma via the parton coalescence
model and from the hadronic matter through hadronic reactions between
nucleons and pions or kaons. We find that a substantial number of $\Theta^+$
are produced from the quark-gluon plasma. Because of the expected narrow
width, the $\Theta^+$ interacts weakly in the hadronic matter. As a result,
the final number of $\Theta^+$ is not much affected by interactions in the
hadronic matter and is thus sensitive to the initial number of $\Theta^+$
produced from the quark-gluon plasma. Study of $\Theta^+$ production in
relativistic heavy ion collision thus provides the possibility of
understanding the dynamics of quark-gluon hadronization.

\section{acknowledgments}

We thank Berndt M\"uller and Jorgen Randrup for comments on an earlier
version of this paper. This Letter was based on work supported in part by the
US National Science Foundation under Grant No. PHY-0098805 and the Welch
Foundation under Grant No. A-1358. VG was further supported by the National
Institute of Nuclear Physics (INFN) in Italy, while LWC by the National
Science Foundation of China under Grant No. 10105008 and SHL by the Korean
Research Foundation under Grant No. KRF-2002-015-CP0074.


\begin{thebibliography}{99}
\bibitem{nakano} {\small T. Nakano et al., LEPS Collaboration, Phys. Rev.
Lett. 91, 012002 (2003).}

\bibitem{stepanyan} {\small S. Stepanyan et al., CLAS Collaboration, Phys.
Rev. Lett. 91, 252001 (2003).}

\bibitem{kubarovsky} {\small V. Kubarovsky et al., CLAS Collaboration, Phys.
Rev. Lett. 92, 032001 (2004).}

\bibitem{barth} {\small J. Barth et al., SAPHIR Collaboration, Phys. Lett. B
572, 127 (2003).}

\bibitem{airapetian} {\small A. Airapetian et al., HERMES Collaboration,
Phys. Lett. B 585, 213 (2004).}

\bibitem{barmin} {\small V.V. Barmin et al., Phys. Atom. Nucl. 66, 1715
(2003); Yad. Fiz. 66, 1763 (2003).}

\bibitem{asratyan} {\small A.E. Asratyan et al., Phys. Atom. Nucl. 67, 682
(2004); Yad. Fiz. 67, 704 (2004).}

\bibitem{diakonov} {\small D. Diakonov, V. Petrov, and M. Poliakov, Z. Phys.
A359, 305 (1997).}

\bibitem{prasz} {\small M. Praszalowicz, Phys. Lett. B 575, 234 (2003).}

\bibitem{polyakov} {\small M.V. Polyakov and A. Rathke, Eur. Phys. J. A 18,
691 (2003).}

\bibitem{walliser} {\small H. Walliser and V.B. Kopeliovich, J. Exp. Theor.
Phys. 97, 433 (2003).}

\bibitem{jennings} {\small B.K. Jennings and K. Maltman, Phys. Rev. D 69,
094020 (2004).}

\bibitem{borisyuk} {\small D. Borisyuk, M. Faber, and A. Kobushkin,
hep-ph/0307370.}

\bibitem{itzhaki} {\small N. Itzhaki, I.R. Klebanov, P. Ouyang, and L.
Rastelli, Nucl. Phys. B 684, 264 (2004).}

\bibitem{riska} {\small Fl. Stancu and D.O. Riska, Phys. Lett. B 575, 242
(2003).}

\bibitem{lipkin} {\small M. Karliner and H.J. Lipkin, hep-ph/0307243.}

\bibitem{jaffe} {\small R.L. Jaffe and F. Wilczek, Phys. Rev. Lett. 91,
232003 (2003).}

\bibitem{hosaka} {\small A. Hosaka, Phys. Lett. B 55, 571 (2003).}

\bibitem{glozman} {\small L.Y. Glozman, Phys. Lett. B 575, 18 (2003).}

\bibitem{zhu} {\small S.L. Zhu, Phys. Rev. Lett. 91, 232002 (2003).}

\bibitem{matheus} {\small R.D. Matheus, F.S. Navarra, M. Nielsen, R.
Rodrigues da Silva, and S.H. Lee, Phys. Lett. B 578, 323 (2004).}

\bibitem{sugiyama} {\small J. Sugiyama, T. Doi, and M. Oka, Phys. Lett. B
581, 167 (2004).}

\bibitem{ciskor} {\small F. Csikor, Z. Fodor, S. D. Katz, and T. G. Kov\'{a}%
cs, JHEP 0311, 070 (2003).}

\bibitem{sasaki} {\small S. Sasaki, hep-lat/0310014.}

\bibitem{liu} {\small W. Liu and C.M. Ko, Phys. Rev. C 68, 045203 (2003).}

\bibitem{liu1} {\small W. Liu, C.M. Ko, and V. Kubarovsky, Phys. Rev. C 69,
025202 (2004); W. Liu and C.M. Ko, ibid., 69, 045204 (2004); nucl-th/0309023.%
}

\bibitem{oh} {\small Y. Oh, H. Kim, and S.H. Lee, Phys. Rev. D 69, 014009
(2004).}

\bibitem{nam} {\small S.I. Nam, A. Hosaka, and H.C. Kim, Phys. Lett. B 579,
43 (2004).}

\bibitem{randrup} {\small J. Randrup, Phys.Rev. C 68, 031903 (2003).}

\bibitem{rafelski} {\small J. Rafelski and B. M\"{u}ller, Phys. Lett. B 101,
111 (1982).}

\bibitem{alcor} {\small T.S. Bir\'{o}, P. L\'{e}vai, and J. Zim\'{a}nyi,
Phys. Lett. B 347, 6 (1995); ibid. 472, 243 (2002); Phys. Rev. C 59, 1574
(1999); T.S. Bir\'{o}, T. Cs\"{o}rg\"{o}, P. L\'{e}vai, and J. Zim\'{a}nyi,
Phys. Lett. B 472, 243 (2002); P. Csizmadia and P. L\'{e}vai, Phys. Rev. C
61, 031903 (2000).}

\bibitem{greco} {\small V. Greco, C.M. Ko, and P. L\'{e}vai, Phys. Rev.
Lett. 90, 202302 (2003); Phys. Rev. C 68, 034904 (2003).}

\bibitem{hwa} {\small R.C. Hwa and C.B. Yang, Phys. Rev. C 67, 034902
(2003); 064902 (2003).}

\bibitem{fries} {\small R.J. Fries, B. M\"{u}ller, C. Nonaka, and S.A. Bass,
Phys. Rev. Lett. 90, 202303 (2003); Phys. Rev. C 68 044902 (2003).}

\bibitem{nonaka} {\small C. Nonaka, M. Asakawa, S.A. Bass, R.J. Fries, and
B. Mueller; nucl-th/0312081.}

\bibitem{bjorken} {\small J.D. Bjorken, Phys. Rev. D 27, 140 (1983).}

\bibitem{ko} {\small C.M. Ko, X.N. Wang, B. Zhang, X.F. Zhang, Phys. Lett. B
444, 237 (1998).}

\bibitem{levai} {\small P. L\'{e}vai and U. Heinz, Phys. Rev. C 57, 1879
(1998).}

\bibitem{adcox} {\small A. Basilevsky et al. (PHENIX Collaboration), Nucl.
Phys. A715, 486c (2003).}

\bibitem{ampt} {\small Z.W. Lin, S. Pal, C.M. Ko, B.A. Li, and B. Zhang,
Phys. Rev. C 64 011902 (2001); B. Zhang, C.M. Ko, B.A. Li, and Z.W. Lin,
ibid. 61, 067901 (2000).}

\bibitem{karsh} {\small F. Karsch, Nucl. Phys. A698, 199 (2002).}

\bibitem{kajantie} {\small K. Kajantie, J. Kapusta, L. McLerran, and A.
Mekjian, Phys. Rev. D 34, 2746 (1986).}

\bibitem{rapp} {\small L. Grandchamp and R. Rapp, Phys. Lett. B 523, 60
(2001); R. Rapp, Phys. Rev. C 63, 054907 (2001).}

\bibitem{carlson} {\small C.E. Carlson, C. D. Crone, H.J. Kwee, and V.
Nazaryan, hep-ph/0312325.}

\bibitem{simon} {\small G.G. Simon et al., Nucl. Phys. A333, 381 (1980).}

\bibitem{letessier} {\small J. Letessier, G. Torrieri, S. Steinke, and J.
Rafelski, Phys. Rev. C 68, 061901(R) (2003).}

\bibitem{star} {\small C. Adler et al. (STAR Collaboration), Phys. Rev.
Lett. 89, 092301 (2002); K. Adcox et al. (PHENIX Collaboration), ibid. 89,
092302 (2002).}

\bibitem{xia} {\small C.M. Ko and L.H. Xia, Phys. Rev.C 38, 179 (1988).}

\bibitem{Ko01} {\small C.M. Ko, V. Koch, Z.W. Lin, K. Redlich, M. Stephanov,
and X.N. Wang, Phys. Rev. Lett. 86, 5438 (2001).}

\bibitem{arndt} {\small R.A. Arndt, I.I Strakovsky, and R.L. Workman, Phys.
Rev. C 68, 042201(R) (2003).}

\bibitem{haiden} {\small J. Haidenbauer and G. Krien, Phys. Rev. C 68,
052201(R) (2003).}

\bibitem{nuss} {\small S. Nussino, hep-ph/0307357.}

\bibitem{cahn} {\small R.N. Cahn and G.H. Trilling, Phys. Rev. D 69, 011501
(2004).}

\bibitem{azimov} {\small R.A. Arndt, Ya.I. Azimov, M.V. Polyakov, I.I.
Strakovsky, and R.L. Workman, Phys. Rev. C 69 035208 (2004).}
\end{thebibliography}
\end{document}